# PECULIARIDADES DE LA ECONOMÍA ISLANDESA EN LOS ALBORES DEL SIGLO XXI

**Autores**:

Inés Martín Santos

Universidad Europea de Madrid

Ángel Rodríguez López

Universidad Europea de Madrid

**Resumen**

Se repasa brevemente la historia y las finanzas islandesas de manera diacrónica. Se presenta a Islandia como bastión del estallido de la crisis financiera internacional que comienza a gestarse a principios del siglo XXI y cuyo origen se hace evidente en la fecha simbólica del año 2008. Se analizan las razones fundamentales de esta crisis, centrandonos en las particularidades de la estructura económica islandesa. Se consideran las diferencias y parecidos de esta situación en relación a algunos otros países en similares circunstancias. Se estudia el caso del banco Icesave. Se considera la repercusión que la crisis experimentada por Islandia tiene en el ámbito internacional, especialmente en los inversores extranjeros y en los conflictos jurídicos surgidos a raíz de las medidas adoptadas por el gobierno islandés para sacar al país de la bancarrota.

**Abstract**

Icelandic history and diachronically finances are briefly reviewed. Iceland is presented as a bastion of the outbreak of the global financial crisis begins to take shape in the early twenty-first century and whose origin is evident in the symbolic date of 2008. The main reasons for this crisis are analyzed, focusing on the particularities of Iceland's economic structure. The differences and similarities of this in relation to some other countries in similar circumstances are considered. Bank Icesave case is studied. The impact of the crises experienced by Iceland has in the international arena, especially foreign investors and legal disputes arising out of actions taken by the Icelandic government to pull the country out of bankruptcy is considered.

**Palabras clave:** Islandia, crisis financiera, banca, Icesave, políticas gubernamentales.

## 1. PRECEDENTES

Para analizar la Economía de un país parece conveniente también estar al tanto de sus raíces históricas, estudiar su entorno natural, comprender las actitudes de sus ciudadanos. Se trata de conocer, en definitiva, su cultura. El hecho de que la crisis económica de este país se haya resuelto por medio de manifestaciones pacíficas de los ciudadanos, capaces incluso de reformas constitucionales, no es algo casual, y entronca con la personalidad del pueblo islandés. Seguramente en otros lugares esta actitud sería impensable.

Los orígenes de Islandia se remontan a los primeros asentamientos vikingos en el siglo IX. La falta de una unidad terrestre y las sucesivas luchas internas propiciaron su pérdida de soberanía y la dependencia primero de Noruega, y luego de Dinamarca desde mediados del siglo XIII hasta finales de la segunda guerra mundial.

Durante este tiempo el islandés ha sido un pueblo pobre y sometido. El emigración masiva a Estados Unidos durante la primera mitad del siglo XX es un aspecto ilustrativo de esta coyuntura que frenaba el desarrollo industrial y que, de paso, fomentaba los movimientos independentistas. La inmigración no es significativa pese a contar con un 6% de personas extranjeras, el motivo principal es la dureza del clima y de la geografía física.

La historia moderna presenta a Islandia como una colonia danesa con un sistema social de raíz casi feudal dependiente de Dinamarca y de la Iglesia Luterana (lo que popularmente se conoce como *el pulpo* u oligarquía de unas 14 familias), propietarias de la mitad del suelo cultivable en la gran isla e islotes colindantes.

Islandia es un país situado en el Atlántico Norte, muy cerca del Ártico, con una superficie de 103.000 km² y una densidad de población menor del 3% por km². Un tercio de la misma vive en la capital Reykjavík. Su superficie terrestre es muy escarpada e infértil. La explotación fundamental es la pesca. El Partido de la Independencia es uno de los más sobresalientes, aunque tras la crisis del 2008 y posteriores elecciones ha sido relevado por la Alianza Socialdemócrata en coalición con otros partidos minoritarios.

La población islandesa en siglos pasados por término medio se movió en torno a los 60.000 habitantes con algunos altibajos debidos a pestes y hambrunas. A mediados del siglo XX Islandia tenía una población de 150.000 personas con un crecimiento pequeño pero regular, estancado entre los años 2008 y 2010.

A principios del siglo XXI Islandia alcanzó los 320 000 habitantes, cifra menor que la de algunas ciudades y provincias españolas, por lo que el caso islandés no es extrapolable a la mayor parte del resto de naciones, e incluso su estructura económica podría estudiarse desde la perspectiva microeconómica. Su población activa rondaba las 170.000 personas. Su Producto Interior Bruto era de 13 000 millones de dólares US, y el desempleo y la deuda pública casi inexistentes.

Así pues, antes de la crisis, Islandia se presenta como un país sin mayores complicaciones, un lugar apacible de personas tolerantes, poco ambiciosas, tan amantes de la literatura como de la independencia económica[1], una república democrática (democracia representativa, como la mayoría, y en parte directa por tradición) defensora del libre comercio con las limitaciones habituales de la mayor parte de los países de este tipo.

Pero esta imagen, que pudiera considerarse idílica, con sanidad y educación gratuitas, cambió a principios del siglo XXI. De un sistema económico fundamentalmente caracterizado por el trueque (recordemos que la banca apareció allí en el siglo XIX) se pasó a una economía financiera volcada a la productividad, y a principios del siglo presente, a la especulación.

---

[1] Su solicitud oficial el 16 de julio del 2009 como miembro de la Unión Europea (en contra de la opinión popular, temerosa de perder su autonomía en materia económica, sobre todo la exclusividad de explotación de los caladeros del mar Ártico) se debió más bien a un lavado de imagen tras la crisis del 2008, al aislacionismo en que se encontraba respecto a otros países nórdicos como Suecia y Noruega, y a la pretensión de mejorar la credibilidad del país en los mercados financieros internacionales.

En los años anteriores a la crisis, la economía islandesa no era tan sólida como parecía. Su moneda era muy inestable, se veía sometida a constantes fluctuaciones, y se producía una falta de control de la inflación [MICHELIS, 2009].

Cuando los bancos estatales se privatizaron, estos comenzaron a dejar atrás los créditos familiares y las pequeñas inversiones seguras, y emprendieron una carrera desenfrenada hacia las inversiones de riesgo por las que llegaron a ofrecer interés en torno al 15%. Ello atrajo no sólo a los ahorradores del propio país sino también sorprendentemente a los inversores extranjeros que invirtieron enormes fortunas en una economía con una productividad limitada por la propia envergadura del país.

Gráfico 1. Endeudamiento de las familias islandesas entre 1980 y 2007 en función de sus ingresos

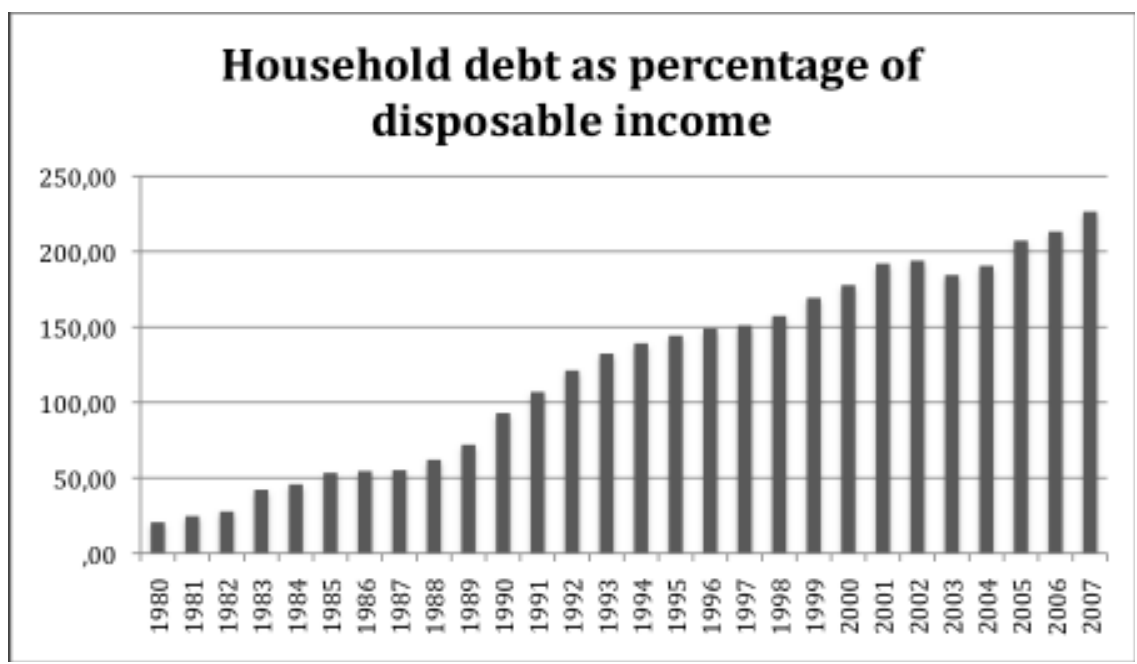

Fuente: Banco Central de Islandia.

Volvamos atrás. Los recursos principales de Islandia antes de la crisis se asentaban en su potente actividad pesquera, su industria manufacturera y de servicios, la producción alumínica, los servicios financieros, el desarrollo del software informático, la biotecnología y el turismo.

En alguna ocasión se ha citado, de manera exagerada, la influencia de Milton Friedman en el gobierno islandés, en 1984, para la aplicación extrema de los principios del liberalismo económico, pero tal consideración no deja de resultar tan anecdótica como la creencia de que la civilización árabe invade en España en el año 911 por la traición del conde don Julián.

En el año 2000 el gobierno islandés implementó una política de desregularización o liberalización económica, tal como anteriormente había aplicado a principios de los años 60 con la reducción de subsidios a la industria pesquera; y también la hizo extensible en los años 80 a la desregularización de los tipos de interés que crecieron por encima de la tasa de inflación [THORVALDUR, 2010, p. 46].

Esta clase de decisiones, como ya advirtió Stiglitz en el año 2001, no dejan de ser peligrosas, al menos cuando se trata de una economía de pequeñas dimensiones difícilmente ajustable a principios generales de política macroeconómica y sujeta a bruscos movimientos especulativos de capital en el entorno internacional.

En estas circunstancias, durante un lustro y medio el precio de la vivienda casi se duplicó (aunque algo menos que en España) como se puede apreciar en el cuadro siguiente:

Gráfico 2. Precio de la vivienda en Islandia.

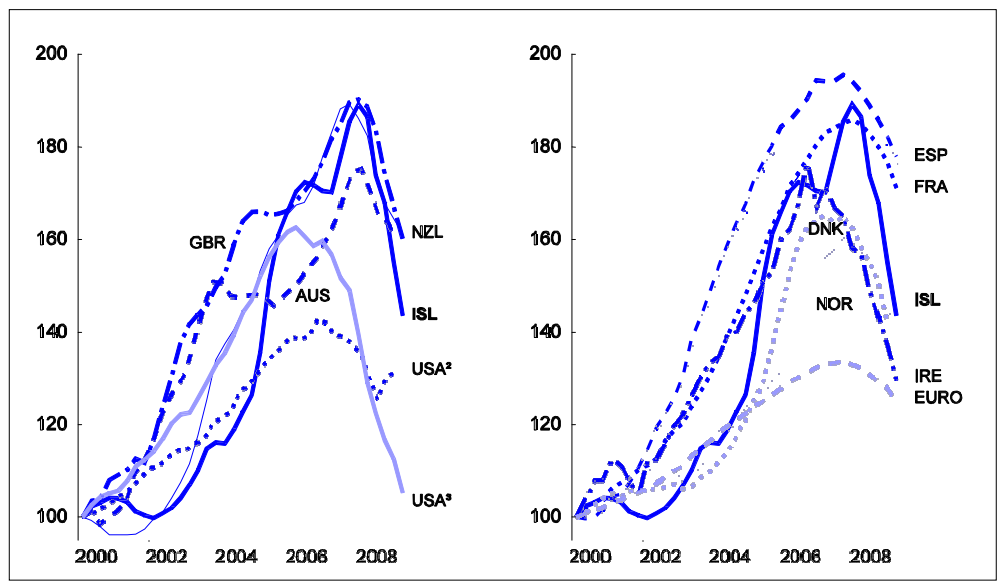

Fuente: Statistics Iceland; Reserve Bank of Australia and OECD Economic Outlook database (apud CAREY, 2009, p. 15).

Islandia pasó a aparecer en el mundo de las finanzas internacionales como un paraíso para la inversión extranjera, opinión respaldada por agencias tan reconocidas como KPMG (acrónimo de los apellidos de sus fundadores) y por informes de expertos de reconocido prestigio como Mishkin y Herberstsson que cobraron 124 000 dólares por un estudio de discutible fiabilidad [2006], publicado el mismo año en que Islandia estaba experimentando una mini-crisis[2].

A ello siguieron otras medidas liberalizadoras como la privatización de los grandes bancos Glitnir, Landsbankinn y Kaupthing en el año 2003, cuyas acciones se revalorización prontamente en un 900%. Durante cinco años recibieron préstamos por valor de $120.000.000.000. Sin embargo, como indica Thorvaldur [2010, p. 46] estos bancos se vendieron por poco dinero a personas con escasa experiencia en el campo de los negocios que, en lugar de realizar inversiones rentables prefirieron recompensar las actividades empresariales de amigos y familiares. A finales de 2008 estos bancos, que habían acumulado unas pérdidas de unos $100.000.000.000, quebraron. La burbuja financiera era previsible. El paro aumentó en el siguiente medio año del 1,5% al 8%.

El deseo de muchos de convertir a Islandia en un enclave financiero internacional, similar a Suiza o Luxemburgo, había comenzado con mal pie, sobre todo si se tiene en cuenta que no era un paraíso fiscal, que sus habitantes son mucho menos negociantes que los suizos y los luxemburgueses, y que sus reservas de divisas, como afirma Matthiasson [2008] no tenían respaldo alguno.

## 2. LAS FINANZAS INTERNACIONALES

Es complicado explicar las actividades económicas nacionales del siglo XXI si no es en el marco internacional, hasta el punto de que podríamos cambiar el término macroeconomía por el de multieconomía.

Uno de los factores que más han influido en el desarrollo de la competencia por todo el mundo ha sido el de la globalización. Sus ventajas y perjuicios en los países en vías de desarrollo han sido tratadas, entre otros, con gran claridad en la parte segunda de la compilación preparada por Eleanor M. Fox y Abel M. Mateus [2011, vol. I, pp. 189-240].

La aldea global, globalización o mundialización, como también se denomina, es un fenómeno que empuja a los gobiernos a la expansión comercial al exterior si pretenden mantener unos mínimos y aceptables niveles de competitividad. El fenómeno de la globalización no es nuevo. Esta, a la que asistimos, se puede considerar como mínimo la tercera globalización[3] (o *mundialización*, como prefieren decir los estudiosos franceses). La crisis internacional que atravesamos tampoco es la única ni probablemente sea la última. En consecuencia deben arbitrarse medidas, desde las instancias políticas, ajustadas a la realidad actual para hacer más viables el desarrollo y las relaciones económicas entre los países.

---

[2] La tasa de crecimiento descendió del 6% al 1%.

La globalización es un fenómeno que empuja a los gobiernos a la expansión comercial al exterior si pretenden mantener unos mínimos y aceptables niveles de competitividad.
Si bien la globalización es un fenómeno prácticamente inevitable en el siglo de la información y de las comunicaciones rápidas y baratas, sin embargo no se han establecido reglas comunes a las que se deban ajustar tanto las empresas multinacionales como internacionales.

A las ya complicadas relaciones económicas internacionales hay que añadir una práctica habitual en el comercio tanto intra como extra-nacional: los derivados, es decir fórmulas de relación indirecta entre las partes principales de un negocio que pueden adoptar formas: venta de la deuda, subcontrataciones,... operaciones que a menudo contribuyen a fomentar la opacidad de los negocios y que el espíritu de libre mercado, en virtud de sus principios, no puede impedir.

La crisis económica del 2008 llegó precedida de una crisis financiera global producida entre 1998 y 1999, cuyas principales víctimas fueron las economías domésticas, y comenzó a anunciarse hacia el año 2006 antes del detonante Lehman Brothers.

El caso islandés no puede entenderse sin analizar los fenómenos de globalización y de competencia que afectan a la mayor parte de los países del globo terrestre. Estas constantes implican dificultades todavía por resolver en muchos aspectos relacionados con la internacionalización de las empresas, con la unificación de criterios de operatividad

---

[3] Previa unificación de la zona mediterránea, la primera se produjo con el descubrimiento de América y el depósito de capitales en centro Europa. La segunda tuvo lugar a raíz de la Revolución Industrial de los siglos XVIII y XIX y el Reino Unido fue el centro internacional de las finanzas. Ver el artículo de Carlos Fuentes. Mundo y localismo: la tercera globalización. *La nación on line*. 30 de noviembre del 2003. Accesible en:
 http://www.almendron.com/politica/pdf/2003/reflexion/reflexion_0170.pdf [Consulta 24 de marzo del 2010]

de los mercados y con la supervisión de los gobiernos sobre este tipo de empresas.

Justamente el fiasco del banco islandés Icesave, como más adelante se verá, ha sido motivo de polémica y lucha de intereses entre los gobiernos de Islandia con los de Holanda y Reino Unido. Pero no sólo fueron holandeses e ingleses los inversores en la economía islandesa, también, aunque no hayan adquirido notoriedad, alemanes y sobre todo norteamericanos.

Este asunto tiene que ver con las decisiones empresariales porque los grandes movimientos de capital no siempre son acertados. Como afirma Stiglizt, "The notion that markets are rational, in some sense, has been called into question" [2001, p. 2].

Además las transacciones económicas internacionales se enfrentan, entre otros escollos, al de las distintas legislaciones nacionales o regionales[4] sobre asuntos económicos e, incluso, a vacíos legales[5].

Al problema de la falta de unificación de las mismas normas para todos los mercados hay que añadir la diversidad cultural de los pueblos y, en consecuencia, el comportamiento general de lo que podemos denominar actitud social como el resultado conjunto de una multiplicidad de idiosincrasias.

En el caso de Islandia no cabe duda de que hubo falta de planificación y equivocaciones que hay que suponer que no fueron intencionadas.

## 3. ANÁLISIS SITUACIONAL

No es oro todo lo que reluce. Desde la segunda guerra mundial Islandia pasó de ser uno de los países europeos más pobres a ocupar el lugar de los más desarrollados en el ranking mundial [MATTHIASSON, 2008, p. 2], de hecho el *Informe sobre el Desarrollo Humano 2007-2008*, publicado por el Programa de las Naciones Unidas para el Desarrollo

---

[4] Se entiende por región en este contexto la amalgama de varios países con características similares.
[5] Un caso llamativo, por ejemplo, es el de la empresa multinacional VISA porque siendo España uno de los países con mayor número de tarjetas de crédito VISA y con un volumen de negocio del 12,4% de las operaciones que realiza en Europa mediante este sistema, sin embargo esta compañía multinacional no paga aquí sus impuestos.

[PNUD, 2008, p. 231] presenta a Islandia como el país con mayor bienestar del mundo.

Pero diversos indicadores económicos ofrecían un aspecto menos favorable como, por ejemplo, el índice de inflación que durante mucho tiempo estuvo oscilando entre el 20 y el 40%, incluso en los últimos años fue muy variable, como se puede ver en el siguiente gráfico:

Gráfico 3. Inflación en Islandia.

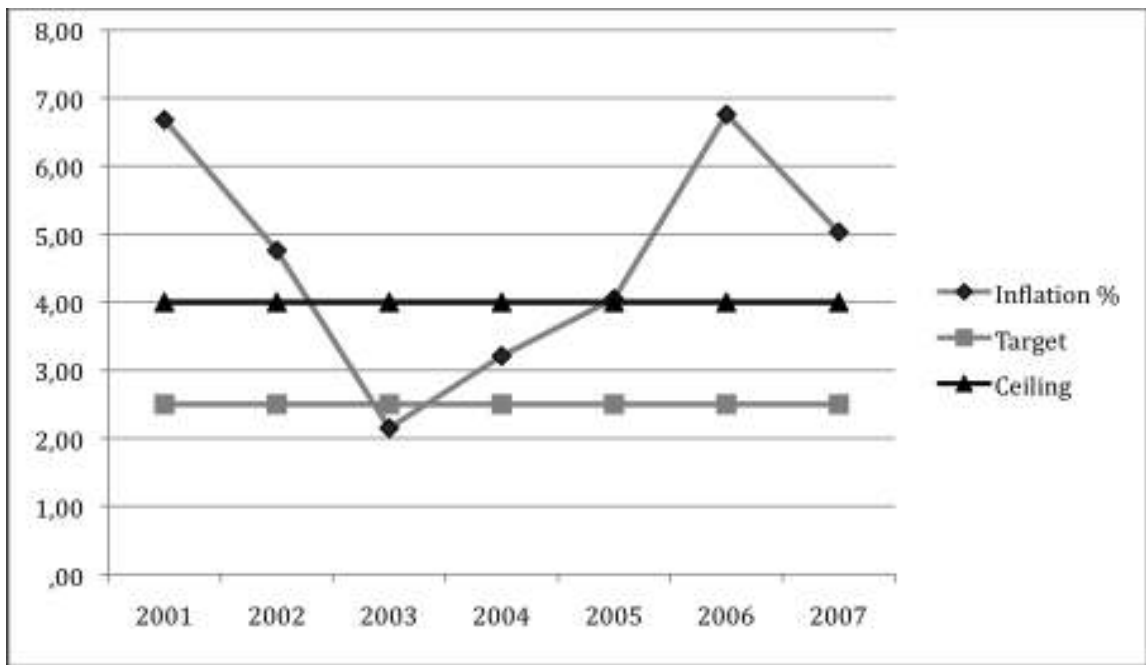

Fuente: *Statistics Iceland*. En: http://www.statice.is/ [Consulta 29/08/2012].

Las transacciones financieras del exterior estaban muy restringidas hasta que Islandia entabló acuerdos con las Unión Europea a principios de 1990.

Tras una larga tradición del trueque, la moneda islandesa fue un elemento débil en el mundo financiero [HERBERTSSON, 2006]. Hasta el año 2000 la corona islandesa tenía una difícil conversión a divisas extranjeras. Y, sin embargo, a pesar de todos estos inconvenientes Islandia se puso de moda para los inversores, en parte motivados por un exceso de confianza en la implementación de medidas liberalizadoras.

Algunos parámetros utilizados por Kristjánsdóttir [2005] como la situación o distancia y el tamaño de mercado, tampoco eran motivos suficientes para atraer la inversión extranjera, al menos la estadounidense.

Las agencias de calificación norteamericanas, probablemente más influidas por la labor de comercialización o márquetin del gobierno islandés que por un concienzudo análisis de la realidad económica, mantuvieron la calificación de sobresaliente para Islandia.

La disyuntiva que a menudo se plantea es si Islandia ha sido una víctima más de la crisis financiera internacional o si simplemente su fracaso económico ha sido fruto de su propia ineficiencia.

Además de consideraciones generales que señalan como causa determinante la excesiva rapidez de los drásticos cambios de producción en la economía islandesa, el sobrecalentamiento económico, la falta de una política económica disciplinada, etc., sin embargo un estudio de Dong-Hyun et al. [2009], tomando en consideración diversos patrones shumpeterianos, demuestra que una parte sustancial de las empresas islandesas se considera no fue eficiente en el sentido de que no usaron la tecnología de la mejor manera posible [p. 19].

Dong-Hyun compara los resultados obtenidos en las empresas manufactureras con las del sector servicios y se encuentra que estas últimas son más eficientes en el uso de las tecnologías que las anteriores.

A esta limitación se une la competencia que tiene que asumir el pequeño tamaño de su mercado interno en relación con el de sus vecinos y el de otros países más lejanos.

La crisis islandesa, en líneas generales, se produjo como la crisis irlandesa: grandes créditos, gran inversión a ciegas e incapacidad para devolver los intereses de la deuda e incluso la misma deuda. Pero la diferencia fundamental fue que mientras Irlanda pertenecía a la Unión Europea y podía utilizar como divisa el euro, Islandia se venía obligada a usar la corona islandesa necesariamente devaluada.

A finales del 2008, Islandia experimentó una de las más acusadas crisis financieras en el mundo desde la Segunda Guerra Mundial [SPURK, 2010, p. 10]. Sin alcanzar las proporciones que hubiera tenido en un país de mayor envergadura como España, sin embargo Islandia es el espejo en el que se puede mirar cualquier economía que haya experimentado o que pueda sufrir un desfalco financiero a principios del siglo XXI.

El cambio drástico de un sistema de producción proteccionista a un sistema liberal nos parece una dicotomía simple y parcial [Cfr. WADE. 2012] porque como tales neoliberales sólo podemos considerar a un reducido grupo de políticos con pocos conocimientos de Economía.

Gran parte de los analistas coinciden en señalar como raíces del problema por un lado la ausencia de una política económica que produjo una crisis bancaria, y por otro lado una crisis ocasionada por la especulación de su divisa, "*twin crisis: A bank-crisis and a currency crisis at the same time*) [HANNIBALSSON. 2009, p.3].

Esta consideración no es del todo cierta porque el Banco Central ya en 2001 había establecido unas *targeting rules* conducentes al control de la tasa de inflación. Entre otras medidas figuraba el establecimiento de los tipos de interés en el 15%.

Es poco científico aludir a fenómenos fantasmagóricos como el club Bilderberg como una especie de mano negra capaz de mover los hilos de la economía mundial a su antojo, pero tampoco hay que menospreciar publicaciones típicas del periodismo científico como el *best-seller* de Matt Taibbi, *Cleptopía* [2011].

Si hay que mencionar responsables se debe hacer con la mayor objetividad posible y para ello las contribuciones a los foros de discusión resultan en muchos casos tendenciosas. Que en Islandia hubo una injerencia externa en su economía, más emocional que racional, es evidente. Pero si partimos de la teoría de Minsky y tenemos en cuenta la aportación de D'Atellis [2008] para las pequeñas economías, la economía islandesa era mucho más propensa a la inestabilidad que otras economías nacionales de mayor envergadura.

Sea como fuere, en setiembre del 2008 la banca islandesa se encontró con pasivos de inversores extranjeros que, amedrentados por el caso Lehman Brothers, se apresuraron a retirar sus fondos. La incapacidad para hacer frente a esta demanda obligó a los banqueros islandeses a solicitar ayuda a sus vecinos nórdicos, a Rusia y finalmente al Fondo Monetario Internacional [SHEETAL, 2009, p. 21].

**4. EL CONFLICTO ICESAVE**

La crisis islandesa no sólo se ha producido por las circunstancias apuntadas anteriormente. No ha sido únicamente una crisis ocasionada por una mala gestión financiera sino también por una deleznable actitud especulativa de la banca.

Aprovechando la predicada envidiable situación islandesa afianza sobre todo en su población capacitada, su gran industria pesquera y la producción y exportación de aluminio; los tres grandes bancos islandeses fomentaron la venta de divisas. Ello ocasionó una crisis añadida [LAKEHAL-AYAT, 2010] como explico a continuación.

La confianza en la corona islandesa aparecía como un refugio seguro para otras monedas. Con una excelente labor de imagen por parte de los banqueros islandeses, muchos inversores extranjeros compraron esta divisa que, además ofrecía unos intereses superiores a la media de otras monedas en sus respectivos países.

Una política de préstamos parecida se practicó con los propios islandeses, y la deuda doméstica cuando se produjo la quiebra bancaria en 2008 por un total de 85.000 millones de dólares, era cuantiosa.

Ante esta adversidad los bancos principales, ya nacionalizados, han ido perdonando a los particulares créditos equivalentes al 13% de su PIB, deudas que superaban en muchos casos el 110% del valor de sus viviendas. Esta clase de decisiones junto con la posibilidad de poder devaluar la corona para mejorar la competitividad, ha hecho que la opinión pública islandesa sea contraria a su completa integración en la Unión Europea.

La corona islandesa no sólo estaba sobrevalorada, sino que los mismos bancos que la habían elevado a las nubes, a su vez intentaban bajarla a los infiernos como de hecho sucedió en 2008 con una devaluación del 50% frente al euro. El banco-enseña fue Icesave.

Icesave, de modo similar a ING u otras entidades financieras fue una organización transnacional. Icesave fue un banco que únicamente operaba en internet, dependiente del banco inicialmente privado islandés Landsbankinn o Landsbanki.

Desde octubre del 2006 funcionó en el Reino Unido, alcanzando unos 300.000 ahorradores, de los que consiguió cuatro mil millones de libras esterlinas en depósitos, a los que abonaba un 6% de interés.

En circunstancias parecidas operó en los Países Bajos a partir de mayo del 2008 con 125.000 clientes de los que obtuvo mil setecientos millones de euros con los que concertó entre un 5% y posteriormente un 5,25% de interés.

En Islandia no había una regulación explícita sobre gestión bancaria, al contrario de lo que establece la European Economic Association (EEA) que estipula un mínimo de 20.887 euros de garantía para los depósitos individuales [HANNIBALSSON, 2009, p.7]. lo que promovió que algunos de sus bancos se adentraran en aventuras financieras con escaso capital y ambiciosas expectativas nebulosas.

Cuando en octubre del 2008 estalló la crisis, el banco Icesave presentaba un déficit de unos tres mil millones de euros. Los gobiernos holandés e inglés asumieron las pérdidas ocasionadas a sus ciudadanos, pero cuando Islandia primero dejó caer Landsbankinn y luego lo nacionalizó surgió la presión para que Islandia asumiera dicha pérdida, puesto que en todo momento Icesave se había presentado como un banco filial de Landsbankinn. Las presiones del Reino Unido y de Holanda ante el Fondo Monetario Internacional para recuperar su pago han sido constantes.

Entre otras medidas, el gobierno británico sacó adelante la *terrorist-law* para tener registradas todas las compañías británicas que estaban en propiedad de Islandia. El gobierno de Islandia y ambos bancos islandeses pasaron a formar parte de la lista negra del gobierno británico junto con otras organizaciones terroristas, entre ellas *al Qaeda*.
Islandia se niega a devolver todo el dinero apostado, y por el momento este conflicto no ha dejado satisfechas a las partes interesadas. Recientemente la Asociación Europea de Libre Comercio se ha pronunciado en contra de las pretensiones británicas y holandesas [MILLÁN, 2013, p. 29].

## 5. LA SALIDA DE LA CRISIS

En general si los bancos tienen problemas, los gobiernos suelen nacionalizarlos, y cuando acaban de reflotarse los privatizan. Probablemente esta no sea una medida justa, pero en el entorno macroeconómico pretende evitar mayores catástrofes. En el siglo XIX español la bancarrota del Estado se solucionó de modo provisional con las desamortizaciones de Madoz y Mendizábal.

Los gobiernos de la Unión Europea han obrado de esta manera como lo muestra el cuadro siguiente:

Gráfico 4. Apoyo de los gobiernos de la zona euro al sector financiero, octubre de 2008 – mayo de 2010 en miles de millones de euros

|  | Comprometidos | Implementados | Implementados/Ejecutados (porcentaje) |
|---|---|---|---|
| Inyecciones de capital | 231 | 84,2 | 36 |
| Garantías del pasivo | 1694 | 506,2 | 30 |
| Protección de activos | 238 | 48,7 | 20 |

Fuente: Stotz y Wedow (2010, cuadro 1, p. 24).

En el caso islandés, cuando se reconoció el colapso económico en octubre del 2008, el gobierno nacionalizó los tres mayores bancos del país y su deuda pública empezó a aumentar.

Los posteriores cambios de poder y referendos ha propiciado en los medios de comunicación de masas un frotarse las manos por el encarcelamiento de algunos de los principales responsables políticos de la crisis islandesa, sin embargo esto no impide que el país deje de saldar su deuda externa.

En 2009 el Parlamento acordó devolver la deuda a Holanda, Reino Unido y a algunos otros ahorradores de Alemania y Estados Unidos, sus principales acreedores.

Cuadradas las cuentas, cada familia islandesa debería pagar unos 3.500 euros durante 15 años al 5,5 % de interés. Ante las quejas populares se convocó un referéndum y se acordó bajar el interés al 3% y aumentar el periodo de carencia a 37 años.

Islandia se vio obligada a solicitar un rescate del FMI para renegociar su deuda externa[6] y acometió duros ajustes económicos, pero lejos de refinanciar los tres grandes bancos nacionales los dejó hundir. Esta decisión es difícilmente aplicable en otros países, entre otras razones porque en este caso la mayoría de los activos financieros de la banca islandesa pertenecían a inversores extranjeros y sobre todo porque tal determinación podría ocasionar revoluciones sociales de trágicas consecuencias.

La recuperación islandesa puede parecer milagrosa, pero ha costado cuantiosas pérdidas para muchos inversores que se han visto sometidos a *quitas*; y se ha conseguido fundamentalmente con la devaluación de la moneda, lo que ha propiciado la recapitalización tanto de las economías domésticas como de las arcas del Estado. Si en el año 2009 el índice de inflación era del 16,3%, dos años después se redujo al 4,2%.

No obstante se ha llegado a poner en duda la recuperación de la economía islandesa. La guerra de cifras que se dan se asienta en función del PIB, el cual ha ido disminuyendo desde el año 2008. La recuperación islandesa se puede decir que es satisfactoria pero relativa mientras no se tenga en cuenta un referente.

La realidad actual presenta un Estado que se ha visto obligado a la refinanciación de la banca en unos porcentajes muy superiores a otros países en similares circunstancias, a una reducción del gasto público que afecta las ayudas sociales en educación y sanidad, a los recortes de las pensiones y al aumento de los impuestos. Alternativamente realiza, como medida provisional, controles a la retirada de capitales [WADE, 2012, p. 138].

Medidas parecidas a las adoptadas por otros gobiernos, la diferencia son los porcentajes. La diferencia consiste en que probablemente los islandeses estén siendo los más ágiles para afrontar sus problemas. Principalmente el impuesto de sociedades (más bajo que en otros países, en torno al 15%) y las enormes inyecciones de dinero a la banca han propiciado nuevamente el crédito que necesitan muchos pequeños empresarios que han abandonado el ámbito de la especulación para dedicarse al campo de la productividad.

Además de los datos apuntados, desde hace décadas la política económica islandesa está tratando de evitar la dependencia del sector pesquero diversificando la productividad y derivándola al terreno de la manufacturación (derivados de las pesca y la pequeña ganadería), de los servicios, de la energía térmica y de la minería (en pequeña escala, ya que principalmente se trata de explotaciones de bauxita).

---

[6] Se estima que los tenedores de esta deuda padecieron una quita del 70% de sus ahorros.

Además su ingreso en la Unión Europea supondría en parte la renuncia a ciertas prácticas que se desarrollan en exclusiva como la pesca de ballenas. En setiembre del 2012 ya el Parlamento Europeo aprobó la prohibición de importar determinadas clases de pescado procedente Islandia por la excesiva sobreexplotación de algunas especies como la caballa [ICELAND, 2012].

Ante las dificultades de aplicar políticas macroeconómicas en países con un PIB muy pequeño, algunos [Cfr. MICHELIS, 2009] han propuesto como una de las soluciones óptimas la inclusión de Islandia en la Unión Europea, pero hay que tener en cuenta que parte de la población se muestra reacia a ello porque en algunos aspectos implicaría pérdida de soberanía, supondría en parte, por ejemplo, la renuncia a ciertas prácticas que se desarrollan en exclusiva como la pesca de ballenas.

Con el fin de prevenir crisis futuras, TChaidze [2008] defiende que los países con economías pequeñas no dependan primordialmente del sector bancario, ni de una moneda interna, ni desarrollen políticas monetarias independientes, alejadas de las observadas en otros países vecinos o de parecidas circunstancias.

## 6. CONCLUSIÓN

Los cambios drásticos de la economía islandesa, producidos en un corto período de tiempo parecen un fenómeno singular poco comparable al resto de la economía mundial. Diversos factores apoyan esta consideración, entre otros: una sociedad pequeña con gran influencia de los criterios mantenidos en relaciones de vecindad entre los miembros de la misma comunidad, proclives a los sistemas de autogestión; una tendencia histórica, al menos desde la perspectiva cultural, a la quietud, a la falta de ambiciones; y un gran peso de las mujeres en la decisiones de política económica.

La crisis económica islandesa del año 2008 aparece en la literatura económica como el chivato que hace saltar la alarma de una crisis financiera internacional de mayor envergadura. El año 2008 es meramente simbólico y arbitrario porque en realidad esta crisis se venía fraguando desde 1990. Tanto la información de los medios de comunicación de masas como los informes estadísticos del Fondo Monetario Internacional, de la Comisión Económica de la Unión Europea y del propio Banco Central de Islandia, los documentos de trabajo de la OCDE, así como los estudios publicados antes y después del 2008 coinciden en señalar una inversión poco planificada, altamente especulativa y cargada de un alto riesgo.

Las razones por las que el capital extranjero eligió este país como centro de inversión en condiciones adversas todavía no han se han esclarecido suficientemente, pese a la existencia de documentos entre representantes del Fondo Monetario Internacional y un representante del Banco Central de Islandia en los que aparece la existencia de desequilibrios de carácter macroeconómico, déficit en su balanza de pagos y bajas tasas de ahorro [ALEXANDER, 2001], pese a algunos errores que puedan producirse en las predicciones como los que detecta Daníelsson [2008] para los períodos 1995-2002 y 2000-2007.

Como responsable principal de esta crisis se ha señalado en primer lugar el capital extranjero, y en segundo lugar los políticos y banqueros islandeses; pero no hay que olvidar que el endeudamiento doméstico o casero alcanzó asimismo unas cifras muy elevadas; desde 1994 el número de cuentas corrientes, tarjetas de débito y de crédito se triplicó, los préstamos llegaron a crecer hasta un 50% en el 2005, los préstamos en muchos momentos también llegaron a duplicar el porcentaje del Producto Interior Bruto [TUPLIP, 2007, p. 4], algo sorprendente en una población considerada a menudo conformista y de alto nivel cultural, lo que obliga en parte a realizar análisis propios del campo de la Psicología Social.

El hecho de que Islandia se convirtiera durante algún tiempo en un país de moda para la inversión extranjera motivado probablemente más por influjos publicitarios que por un análisis profundo de su potencial económico es una muestra de la tendencia extendida a principios del siglo XXI a las inversiones especulativas a corto plazo[7].

---

[7] Este fenómeno se observa también, por ejemplo, en la bolsa española, donde a veces se maneja un volumen de títulos mayor al existente, puesto que en el mismo día las acciones pueden pasar hasta por dos, tres o cuatro manos.

Aunque la división entre política y economía sea difícil de desligar, no es incuestionable, y parece aconsejable utilizar criterios distintos para solventar los problemas sociales. Establecer medidas económicas motivadas por actitudes políticas no parece el mejor sistema de arreglar los desajustes sociales, menos aun cuando hace ya más de medio siglo se ha postulado el fin de las ideologías [BELL, 1960].

Dada la gravedad de la crisis económica actual, no sólo en el ámbito nacional sino internacional, parece necesario hacer una llamada a los responsables máximos de la economía mundial para que se establezcan unas mínimas medidas intervencionistas que no condenen a la pobreza a millones de personas. En mi opinión, hoy día el sistema económico se asienta más en ejercicios especulativos que en actividades productivas.

Retomando unas cifras harto significativas que ofrece John Eatwel [apud Chomsky, 1997, p. 74], "en 1970, cerca de 90% del capital internacional se destinaba al comercio y a la inversión de largo plazo –actividades más o menos productivas- y 10% a la especulación. En 1990, las cifras se habían invertido: 90% correspondía a especulación y 10% a comercio e inversión de largo plazo".

La solución islandesa a la crisis provocada por la ambición humana incontrolada de unos pocos ha marcado alguna distancia respecto a circunstancias parecidas en otros lugares. Dista de las soluciones adoptadas en algunos otros países que han sufrido una crisis parecida, en el sentido de que se han reclamado responsabilidades personales y se ha procesado a los responsables que ocultaron el estado de las cuentas al gobierno. Pero desconocemos sin aquí están todos los que son.

Esta conducta no es de extrañar porque la cultura islandesa nunca fue partidaria de depositar confianza ciega en líderes o dirigentes sino más bien en adoptar decisiones de carácter asambleario. Esto que se podría considerar débil afianzamiento de un Estado nacionalista es lo que quizá llevara a este país a una dependencia de Noruega y Dinamarca durante más de diez siglos.

Con todo, el comportamiento del gobierno islandés ha sido legal (no sabemos si justo), es decir, pagar la mayor parte de su deuda porque Islandia no puede convertirse en una isla mayor de la que es y escapar al entramado de la economía mundial. El desastre islandés puede producirse en cualquier otra economía nacional y obligar a pensar, como hace

Sheetal si debe la población asumir los errores de sus dirigentes y apostilla: "Most civilised societies limit punishment to the perpetrator of the deed, even when the stakes involved are very high" [2009, p.39].

A modo de colofón, podemos aseverar que la tendencia del capital de carácter meramente especulativo viene siendo influir en políticas económicas deflacionistas, lo que ocasiona no sólo baja inflación sino también bajo crecimiento y bajos salarios.

**BIBLIOGRAFÍA**